\documentclass[usenatbib,times,usegraphicx,useAMS]{mn2e}
\usepackage{amssymb,url,times}

{\left\lbrace\begin{array}{@{}l@{}}}%
{\end{array}\right.}

\defcitealias{CM04}{CM}
\defcitealias{facchini13_1}{FL}

\title[Wave-like warp propagation in circumbinary discs II]{Wave-like warp propagation in circumbinary discs II.\\ Application to KH 15D}

\author[Lodato \& Facchini]{Giuseppe Lodato$^{1}$, Stefano Facchini$^{1,2}$\thanks{facchini@ast.cam.ac.uk}\\
$^1$Dipartimento di Fisica, Universit\`a Degli Studi di Milano, Via Celoria, 16, Milano, 20133, Italy\\
$^2$Institute of Astronomy, Madingley Road, Cambridge CB3 OHA\\
}
\date{Submission date}
\pagerange{\pageref{firstpage}--\pageref{lastpage}} \pubyear{2013}

\begin{document}
\label{firstpage}
\bibliographystyle{mn2e}
\maketitle

\begin{abstract}

KH 15D is a protostellar binary system that shows a peculiar light curve. In order to model it, a narrow circumbinary precessing disc has been invoked, but a proper dynamical model has never been developed. In this paper, we analytically address the issue of whether such a disc can rigidly precess around KH 15D, and we relate the precessional period to the main parameters of the system. Then, we simulate the disc's dynamics by using a 1D model developed in a companion paper \citep{facchini13_1}, such that the warp propagates into the disc as a bending wave, which is expected to be the case for protostellar discs. The validity of such an approach has been confirmed by comparing its results with full 3D SPH simulations on extended discs. In the present case, we use this 1D code to model the propagation of the warp in a narrow disc. If the inner truncation radius of the disc is set by the binary tidal torques at $\sim 1$ AU, we find that the disc should extend out to 6-10 AU (depending on the models), and is therefore wider than previously suggested. Our simulations show that such a disc does reach an almost steady state, and then precesses as a rigid body. The disc displays a very small warp, with a tilt inclination that increases with radius in order to keep the disc in equilibrium against the binary torque. However, for such wider discs, the presence of viscosity leads to a secular decay of the tilt on a timescale of $\approx 3000 (\alpha/0.05)^{-1}$ years, where $\alpha$ is the disc viscosity parameter. The presence of a third body (such as a planet), orbiting at roughly 10 AU might simultaneously explain the outer truncation of the disc and the maintenance of the tilt for a prolonged time.

\end{abstract}



\section{Introduction}
\label{sec:intro}

It is well known that young protostars are commonly accompanied by a circumstellar disc accreting onto the central star. Moreover, the majority of stars form in binary or higher multiple systems \citep{ghez93,simon95,clarke00}, and circumstellar and/or circumbinary discs are likely to be present in these systems \citep{dutrey94,beust_dutrey05}.

Close encounters in the early stages of formation when stars are surrounded by massive discs lead to substantial disc truncation and the tidal effects of such encounters can also lead the outer parts of the disc to become strongly misaligned or warped \citep{heller95,moeckel06}. \citet{bate10} have studied numerically the evolution of such a chaotic environment in star forming regions. An observational signature of these dynamical effects, where a perturbation induces either a warp or a strong misalignment in a circumstellar disc, is the misalignment between the stellar rotation axis (determined mostly by the angular momentum of material accreted early in the star formation process) and the disc rotation axis (which might correspond to the angular momentum of material accreted later, and eventually of planets). In the case of binary systems, an analogous signature is the misalignment between the binary plane and the disc average plane. Moreover, it has been shown that even simple initial conditions with small asymmetries of the original cloud core can produce a misalignment between the two planes \citep{bonnell92}. Such a disc would be subject to the external torque caused by the binary potential, which would generate a warp in the disc itself. Thus, misaligned circumbinary discs are quite likely to occur in young binary systems \citep{CM04,akeson07}.

In a companion paper \citep[][hereafter \citetalias{facchini13_1}]{facchini13_1} we have developed a 1D numerical model describing the temporal evolution of the shape of a generic circumbinary disc. We have focused on the wavelike regime (see section \ref{sec:warp}), which occurs whenever $\alpha < H/R$ \citep{papaloizou_pringle83,papaloizou_lin95}, where $\alpha$ is the standard viscosity parameter \citep{shakura73} and $H/R$ is the aspect ratio of the disc. Instead, when $H/R<\alpha<1$ the equations describing the evolution are diffusive \citep{pringle92}. The bending waves regime is therefore more suitable to describe protostellar discs, where the aspect ratio is likely to be $\sim 0.1$ \citep[e.g. see][]{lodato08}, while $\alpha \leq 0.01-0.1$. In \citetalias{facchini13_1} we have tested our model by comparing our results with full 3D hydrodynamical simulations, and found a very good agreement. 

In this paper, we apply the model developed by \citetalias{facchini13_1} to a specific case: KH 15D. \citet{CM04} (hereafter \citetalias{CM04}) have described such a system as a misaligned dust laden circumbinary ring precessing around a binary system formed by two similar protostars. Such an environment is suitable in order to describe the dynamical evolution of the precessing narrow disc with our model. We will need to make several approximations and assumptions, but we will show that we are able to fully reproduce the precession of the ring, and to make many relevant dynamical considerations.

In section \ref{sec:obs} we summarise the observations of KH 15D and we describe the model derived by \citetalias{CM04}. In section \ref{sec:prec} we derive the theoretical precessional period from our 1D model and we explore the issue of rigid precession. In section \ref{sec:res} we report our results and we compare them with the model by \citetalias{CM04}. Finally, in section \ref{sec:concl} we draw our conclusions.

\section{Modelling the main system parameters}
\label{sec:obs}

KH 15D is an object that presents a T-Tauri like spectrum, and has been classified as a K pre-main-sequence star with an age of $\sim 3$ Myr \citep{herbst2010}. Since 1998, astronomers realised that this object presented a peculiar light curve \citep{kearns1998}, thus this star has been the object of many observations over the years. Moreover, researchers have found archival data of its light curve since 1913. This fact has given the possibility to analyse the temporal evolution of the light curve for almost 100 years. The light curve shows some very interesting features, that are reported in many papers of the last decade \citep{herbst2002,winn2003,johnson_winn2004,CM04,herbst2010,capelo12}. In order to have a detailed description, which is far from being the aim of this paper, interested readers can look at \citetalias{CM04} and \citet{herbst2010}. KH 15D undergoes very deep periodic eclipses ($P=48.35$ days, $\Delta L=3.5$ mag) and duration ($24$ days). This long duration of the minimum light suggests that this cannot be an ordinary eclipsing  binary. Moreover, as reported above, the light curve undergoes a temporal evolution on a longer timescale than the light curve itself. In particular, the eclipsing duty cycle has been increasing, and the in-eclipse light curve has shown a central reversal in brightness, which has lessened in time in terms of amplitude. Finally, from 1913 to 1951, no eclipse was observed. A model with the aim to describe such a system needs to cover all these photometric features.

A relevant question is whether KH 15D is a single or multiple stellar system. \citet{herbst2002} and later \citet{johnson2004} looked for evidence of a companion star via radial velocity measurements, by using highly resolved spectra from VLT and the HIRES spectrometer of Keck I Telescope, respectively. They both found a periodicity in the radial velocity that was compatible with the photometric one, thus pointing that the binarity of the system is one of the key features to model the light curve and its evolution. Thus, radial velocity measurements suggested that KH 15D does have a companion, with almost equal mass ($\sim 0.5M_\odot$) and temperature. Finally, \citet{johnson2004} constrained the orbital eccentricity of the binary to $0.68 < e < 0.80$.

\citet{agol2004} described a model to reproduce KH 15D photometric variability by introducing a warped circumstellar disc. However, these models succeeded in reproducing the light curve of the system over short timescales, but they could not explain its secular evolution.
In this paper we focus on the model derived by \citet{winn2004} and \citetalias{CM04}, where the photometric variability is caused by the partial occultation of the eccentric binary system by a narrow ring. This ring precesses rigidly on a plane that is inclined by an angle $\bar{I}>0$ with respect to the binary one, and eclipses occur whenever its ascending or descending node regresses into our line of sight toward the two stars. The observer is assumed to look at the binary plane nearly edge-on. For the geometrical details, the interested reader can look at figs. 1 and 2 of \citetalias{CM04}. For our purposes, we will not discuss the uncertainties of this model in great detail, and will simply assume their derived parameters. In particular, we denote the star KH 15D with K, and its orbital companion with K'. The semimajor axis with respect to the centre of mass of the two stars, which have an equal mass $M_K=M_{K'}=0.5M_{\odot}$, is $a_K=a_{K'}=0.13$ AU, so that the separation is $a=0.26$ AU. In order the light curve to be explained, a precession period of $\sim 2770$ yr is required. Obviously, from the measured light curve, the estimate of the precessional period is approximate, and there might be some degeneracy with other parameters (such as the disc radial extent $\Delta R$, or its mean inclination $\bar{I}$). 

Given the model above, we are interested in the dynamics of the rigidly precessing ring, and we want to address the following questions:
\begin{enumerate}
\item Is this precessing ring stable?
\item  Can the ring maintain rigid precession? How large need the warp be in order to prevent differential precession? 
\item Can such a ring stay misaligned with respect to the binary plane over a reasonable timescale?
\end{enumerate}

\section{Precessional period and rigid precession}
\label{sec:prec}

\subsection{Relating the precession frequency to the forcing potential}

The model of section \ref{sec:obs} considers a narrow disc rigidly precessing around the two binary stars. First of all, we relate the precessional frequency to the forcing torque generated by the spherical asymmetry of the system. We know that ${\bf T}(R)={\bf \Omega}_{\mathrm{ext}}(R)\times {\bf L}(R)$ \citep{lodato_pringle06, facchini13_1}, where ${\bf \Omega}_{\mathrm{ext}}(R)$ is the free precession induced at a given radius $R$ by the binary, ${\bf T}(R)$ is the external torque density and ${\bf L}(R)$ the angular momentum of the disc per unit area. We recall that ${\bf L}=\Sigma R^2\Omega {\bf l}$, where $\Sigma$ defines the surface density, $\Omega$ is the angular frequency and ${\bf l}(R)$ is the unit vector indicating the local direction of the angular momentum. If we suppose that the vector ${\bf \Omega}_{\mathrm{ext}}$ is along the $z$-direction (perpendicular to the binary plane), the only interesting components of ${\bf L}$ and ${\bf T}$ are the ones along the $x$ and $y$-axes. Therefore we can rewrite the above relation in terms of a scalar equation:
\begin{equation}
\label{eq:kh15d_tr}
T(R)=\Omega_{\rm ext}(R)L(R),
\end{equation}
where $L(R)=\Sigma R^2\Omega\sqrt{l_x^2+l_y^2}$ is the modulus of the projection of ${\bf L}$ on the $xy$ plane. The variable $\Omega_{\mathrm{ext}}(R)$ is the precessional frequency at which an isolated annulus would precess if it were subject to the external torque $T(R)$. We now suppose that the whole disc precesses with a global precessional frequency $\Omega_{\rm p}$. By requiring that $\Omega_{\rm p}$ is the same for the whole disc (rigid precession) and by integrating equation \ref{eq:kh15d_tr} from $R_{\mathrm{in}}$ to $R_{\mathrm{out}}$ (defined as the inner and the outer edge of the disc, respectively) we obtain:
\begin{equation}
T_{\mathrm{tot}}=\Omega_{\rm p}L_{\mathrm{tot}},
\end{equation}
where we have defined
\begin{equation}
T_{\mathrm{tot}}\simeq\int_{R_{\mathrm{in}}}^{R_{\mathrm{out}}}{\Omega_{\mathrm{ext}}(R)L(R)2\pi R\mbox{d}R},
\end{equation}
and
\begin{equation}
L_{\mathrm{tot}}\simeq\int_{R_{\mathrm{in}}}^{R_{\mathrm{out}}}{L(R)2\pi R\mbox{d}R},
\end{equation}
that are appropriate for small amplitude warps (which are expected for our ring, see below). Thus, we obtain the following relation for $\Omega_{\rm p}$:
\begin{equation}
\label{eq:precession}
\Omega_{\rm p}=\frac{\int_{R_{\mathrm{in}}}^{R_{\mathrm{out}}}{\Omega_{\mathrm{ext}}(R)L(R)2\pi R\mbox{d}R}} {\int_{R_{\mathrm{in}}}^{R_{\mathrm{out}}}{L(R)2\pi R\mbox{d}R}}.
\end{equation}
We then use the following relations:
\begin{equation}
\label{eq:parametrization_precession}
\Omega_{\mathrm{ext}}(R)\propto R^{-s},\ \ \ \ \ \ \ L(R)=\Sigma R^2\Omega\sqrt{l_x^2+l_y^2}\propto R^{1/2-p},
\end{equation}
where we also assume that $\Sigma\propto R^{-p}$. In the second relation we have considered $\sqrt{l_x^2+l_y^2}\approx\mathrm{const.}$ (that is, a small warp), a condition that is usually largely verified, especially in the case of KH 15D (see below). For an external torque due to a central binary, which is the case of KH 15D, the parameter $s=7/2$. We have kept $s$ as a generally free parameter because the same relation could also describe a disc precessing around a spinning black hole. In that case, $s$ would be equal to $3$. We scale the radial variable to the inner disc radius and define $x=R/R_{\mathrm{in}}$. By implementing the relations \ref{eq:parametrization_precession} in equation \ref{eq:precession} we obtain:
\begin{equation}
\Omega_{\rm p}=\Omega_{\mathrm{ext}}(R_{\mathrm{in}})\frac{\int_1^{x_{\rm out}}{x^{3/2-p-s}\mbox{d}x}} {\int_1^{x_{\rm out}}{x^{3/2-p}\mbox{d}x}},
\end{equation}
where $x_{\rm out}=R_{\rm out}/R_{\rm in}$. In reasonable setups, the condition $0<p<5/2$ is verified, and since we know that $s\geq3$, we have that $s+p>5/2$. Therefore we obtain:
\begin{equation}\label{eq:prec_1}
\Omega_{\rm p}=\Omega_{\mathrm{ext}}(R_{\mathrm{in}})\frac{1-x_{\rm out}^{5/2-p-s}} {x_{\rm out}^{5/2-p}-1}\frac{(5/2-p)}{(s+p-5/2)}.
\end{equation}
In the case where $R_{\rm out}/R_{\mathrm{in}}\gg1$, by knowing that the last factor on the r.h.s., that depends on $s$ and $p$, is $\approx O(1)$, we can approximate relation \ref{eq:prec_1} to the following one:
\begin{equation}
\label{eq:t_p_r}
\Omega_{\rm p} \approx \Omega_{\mathrm{ext}}(R_{\mathrm{in}})\left(\frac{R_{\rm out}}{R_{\mathrm{in}}}\right)^{p-\frac{5}{2}}.
\end{equation}
Equation \ref{eq:t_p_r} contains two factors. The first indicates that the precession is forced by the external torque at the inner edge. The second is due to the fact that most of the angular momentum lies in the outer region of the disc (since $M(R)\propto R^{2-p}$), therefore the larger the outer radius, the longer the precessional period.

In order to have an estimate of the free precessional frequency, we need the forcing potential of the binary case. Therefore we focus on the case where $s=7/2$. In \citetalias{facchini13_1} we have deduced it as a time independent gravitational potential, in which the forcing term depends on the non spherical symmetry of the gravitational source. The approximation is to spread the mass of the two stars on two circular orbits, around their centre of mass, and to determine the potential up to the second order in terms of $a/z$ and $z/R$. Note that the strong assumption in this case is to consider the two orbits as circular, when we know that the eccentricity is quite high. We evaluate the external torque by using the expression:

\begin{equation}
\label{eq:external_torque}
{\bf T}=-\Sigma R^2\Omega\left(\frac{\Omega_z^2-\Omega^2}{\Omega^2}\right)\frac{\Omega}{2}{\bf e}_z\times{\bf l},
\end{equation}
where ${\bf e}_z$ is the unit vector perpendicular to the binary orbit and $\Omega_z$ the vertical oscillation frequency, which is related to the gravitational potential \citep[]{lubow_ogilvie2000,lubow_ogilvie01}. Defining the total mass $M=M_K+M_{K'}$, the factor $\eta=M_KM_{K'}/M^2$ and the binary separation $a=a_K+a_{K'}$, the expression for the precession frequency is given by:

\begin{equation}
\label{eq_kh15d_ext_torque}
\Omega_{\mathrm{ext}}(R)=\frac{3}{4}\frac{\eta a^2}{R^2}\left(\frac{GM}{R^3}\right)^{1/2},
\end{equation}
By using this torque to estimate $\Omega_{\mathrm{ext}}(R_{\mathrm{in}})$ in equation \ref{eq:prec_1}, we obtain:
\begin{equation}
\label{eq:kh15d_omega_p}
\nonumber \Omega_{\rm p} = \frac{3}{4}\left(\frac{5/2-p}{p+1}\right)\frac{\eta a^2}{R_{\rm in}^2}  \frac{(1-x_{\mathrm{out}}^{-(p+1)})}{(x_{\mathrm{out}}^{5/2-p}-1)} \Omega_{\rm in}
\end{equation}
where $\Omega_{\rm in}^2=GM/R_{\rm in}^3$ (cf. the analogous expression in equation 1 of CM). Note that this is independent of other relevant disc parameters (for example, the scale height $H_{\rm in}/R_{\rm in}$ or the viscous parameter $\alpha$, defined via the usual \citet{shakura73} prescription). This derivation is similar to the one performed by \citet{bate00}.

\subsection{Conditions to maintain rigid precession}

Hitherto we have deduced the precessional frequency of the disc by assuming that it can precess as a rigid body. In order to do so, we have mostly used the parameters defining the gravitational potential. Now we address the issue of whether a disc, and in particular KH 15D, can precess rigidly. Therefore now we focus on the  properties of the disc. We have already defined the surface density profile.  We then know that the sound speed is related to the disc scale height by $H=c_{\rm s}/\Omega$, and we will assume that $c_{\rm s}\propto R^{-3/4}$ throughout the paper. We set $H_{\rm in}/R_{\rm in}=0.1$ and $\alpha=0.05$.

\citet{pap_terquem95} and \citet{larwood_pap97} developed the arguments introduced by \citet{papaloizou_lin95}, and deduced that the disc can rigidly precess when:

\begin{equation}
\label{eq:rigid}
\Omega_{\rm p} < \frac{c_{\mathrm{s}}} {\Delta R},
\end{equation}
where $\Delta R=(R_{\mathrm{out}}-R_{\mathrm{in}})$. The sound crossing time for the disc thus needs to be shorter than the inverse of the precession frequency. In other words, in order to precess rigidly, the disc needs to communicate the precessional frequency to the outer radii, since the precessional frequency is mostly fixed by the external torque at the inner edge where the external torque is stronger. From equation \ref{eq:rigid} we obtain:

\begin{equation}\label{eq_kh15d_condition1}
\frac{\Delta R}{R}<\frac{H}{R}\frac{\Omega}{\Omega_{\rm p}},
\end{equation}
which evaluated at the outer edge becomes, in a slightly different form:
\begin{equation}
\label{eq:condition}
\frac{\Delta R}{R_{\mathrm{out}}} < \frac{H_{\mathrm{out}}}{R_{\mathrm{out}}} \frac{\Omega_{\mathrm{out}}}{\Omega_{\rm p}}.
\end{equation}
There is one more requirement that the disc has to verify. In the presence of viscosity, the disc will secularly align with the binary plane. We thus have to require that such alignment time is longer than the precessional period $t_{\rm align}>1/\Omega_{\rm p}$, as well as the time since the tilt was initiated. We briefly address this issue in section \ref{sec:warp}, where we summarise the equations that describe the bending waves propagation in a disc, and that intrinsically regulate the dissipation timescale of the misalignment. 

\subsection{Equations of warp propagation}
\label{sec:warp}

The dynamics of warped discs has been discussed in several papers. When $H/R<\alpha<1$, the evolution of the warp is described by diffusive equations \citep{papaloizou_pringle83,pringle92,ogilvie99}, whereas when $\alpha<H/R$ the evolution is tracked by wave equations \citep{papaloizou_lin95,lubow_ogilvie2000,lubow_ogilvie01}. As discussed in section \ref{sec:intro}, protostellar systems are more likely to be in the bending wave regime, since usually $\alpha<H/R$. In this paper, as we have done in \citetalias{facchini13_1}, we use the formulation given by \citet{lubow_ogilvie2000}. Under the assumption that $\partial_t \Sigma\approx 0$, the linearised equations are:
\begin{equation}
\label{eq:wave_l_real}
\Sigma R^2\Omega\frac{\partial {\bf l}}{\partial t}=\frac{1}{R}\frac{\partial {\bf G}}{\partial R}+{\bf T},
\end{equation}
and
\begin{equation}
\label{eq:wave_g_real}
\frac{\partial{\bf G}}{\partial t}+\left(\frac{\kappa^2-\Omega^2}{\Omega^2}\right)\frac{\Omega}{2}{\bf e}_z\times{\bf G} +\alpha\Omega{\bf G}=\Sigma R^3\Omega\frac{c_{\mathrm{s}}^2}{4}\frac{\partial{\bf l}}{\partial R},
\end{equation}
where ${\bf T}$ is the external torque expressed in equation \ref{eq:external_torque}, $\Omega_z$ is the vertical oscillation frequency, $\kappa$ is the epicyclic frequency and $2\pi{\bf G}$ is the internal torque. The explicit expressions for the relevant frequencies $\Omega$, $\Omega_z$ and $\kappa$ can be found, e.g., in  \citepalias[see][]{facchini13_1}. Note that the third term of the r.h.s. of equation \ref{eq:wave_g_real} dissipates the waves through an exponential factor \citep{bate00}. This term indicates that for linear bending waves the damping occurs on a timescale $t_{\rm damp}\approx 1/(\alpha\Omega)$. Even in the case in which the disc precesses as a rigid body, and the tilt reaches a steady state solution (a stationary wave), viscosity still dissipates energy, and the disc tends to get to the lower energy state, which is the alignment with respect to the binary plane. For such a rigidly precessing disc, \citet{pap_terquem95}, \citet{larwood96} and \citet{bate00} estimate that the alignment timescale $t_{\rm align}$ is of the order of the viscous time scale $t_{\nu}=R_{\rm out}^2/\nu$, where $\nu$ is the kinematic viscosity ($\nu=\alpha c_{\rm s}H$). More precisely, \citet{bate00} \citep[see also][]{lubow_ogilvie01} derived an expression for $t_{\rm align}$:

\begin{equation}
\label{eq:kh15d_align_time}
t_{\mathrm{align}}=\frac{1}{\alpha\Omega_{\rm p}}\left(\frac{H_{\mathrm{in}}}{R_{\mathrm{in}}}\right)^{2}\left(\frac{\Omega(R_{\mathrm{out}})}{\Omega_{\rm p}}\right).
\end{equation}
This expression has been derived for the cases where $R_{\rm out}/R_{\rm in}\gg 1$. Note that in equations \ref{eq:wave_l_real} - \ref{eq:wave_g_real} dissipation, and therefore alignment, is automatically implemented via the viscous term.


With these equations we have described the disc as a series of annuli that can interact by pressure and viscous forces, and can both evolve in the inclination angle and precess. We can solve them by using the ring code described in \citetalias{facchini13_1}, which is a similar version of the 1D code used by \citet{LOP02} to simulate a disc affected by a Lense-Thirring external torque.

\section{Results}
\label{sec:res}

The discussion we have gone through so far is quite general. In this section we discuss more specifically the case of KH 15D, and compare our results with the ones suggested by \citetalias{CM04}. 

\subsection{Analytical results}
\label{sec:analyt}

By using equation \ref{eq:kh15d_omega_p}, and implementing KH 15D's parameter in it, we obtain an estimate for the precessional period:
\begin{equation}
\label{eq:precper}
T_p = 78.9 \left(\frac{p+1}{5/2-p}\right) \frac {(x_{\mathrm{out}}^{5/2-p}-1)} {(1-x_{\mathrm{out}}^{-p-1})} \ \mathrm{yr}, 
\end{equation}
where we estimate $R_{\rm in}$ as the radius at which the binary potential tidally truncates the disc. From \citet{art_lubow94} we know that for binary systems with such a high eccentricity $R_{\rm in}\approx 4a$. Since $a\approx0.25$ AU, we set $R_{\rm in}=1$ AU. Note that this value is much lower than the one assigned by \citetalias{CM04} in their work. We can now derive an estimate for $R_{\rm out}$ from equation \ref{eq:precper}, by setting $T_p=2770$ yr. We set the parameter $p$ to two possible values: $p=0.5$ and $p=1$. We obtain $R_{\rm out}=6.71$ and $9.00$ AU, respectively (see Table \ref{tab:kh15d}). Both values are larger than the one used by \citetalias{CM04} in their numerical model. 

More important, we have checked that the disc verifies the condition expressed by relation \ref{eq:condition}, with the parameters of both models. Such discs can therefore precess rigidly.

\subsection{Numerical results}
\label{sec:numer}

\begin{figure*}
\begin{center}
\includegraphics[width=.8\columnwidth]{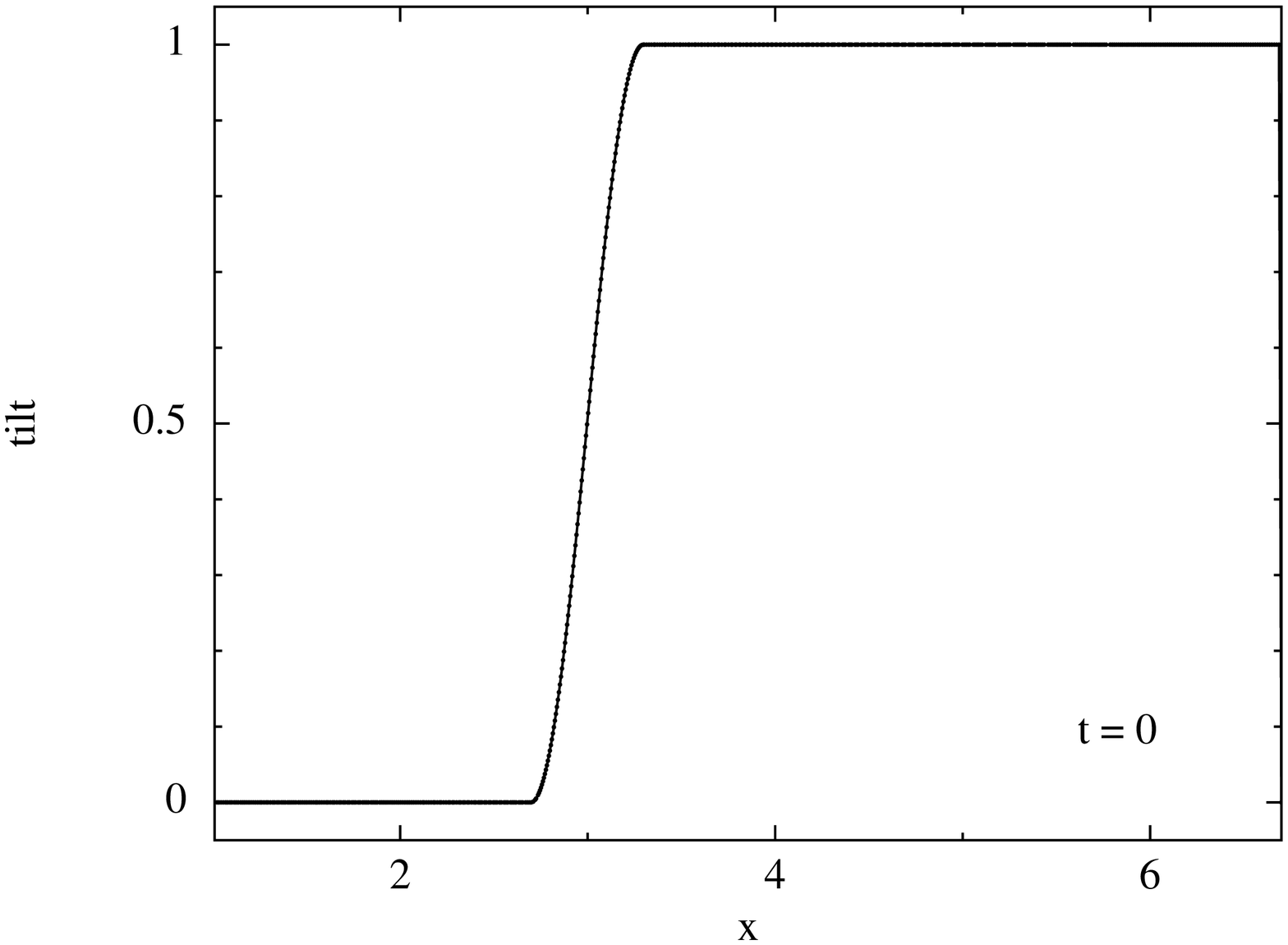}
\includegraphics[width=.8\columnwidth]{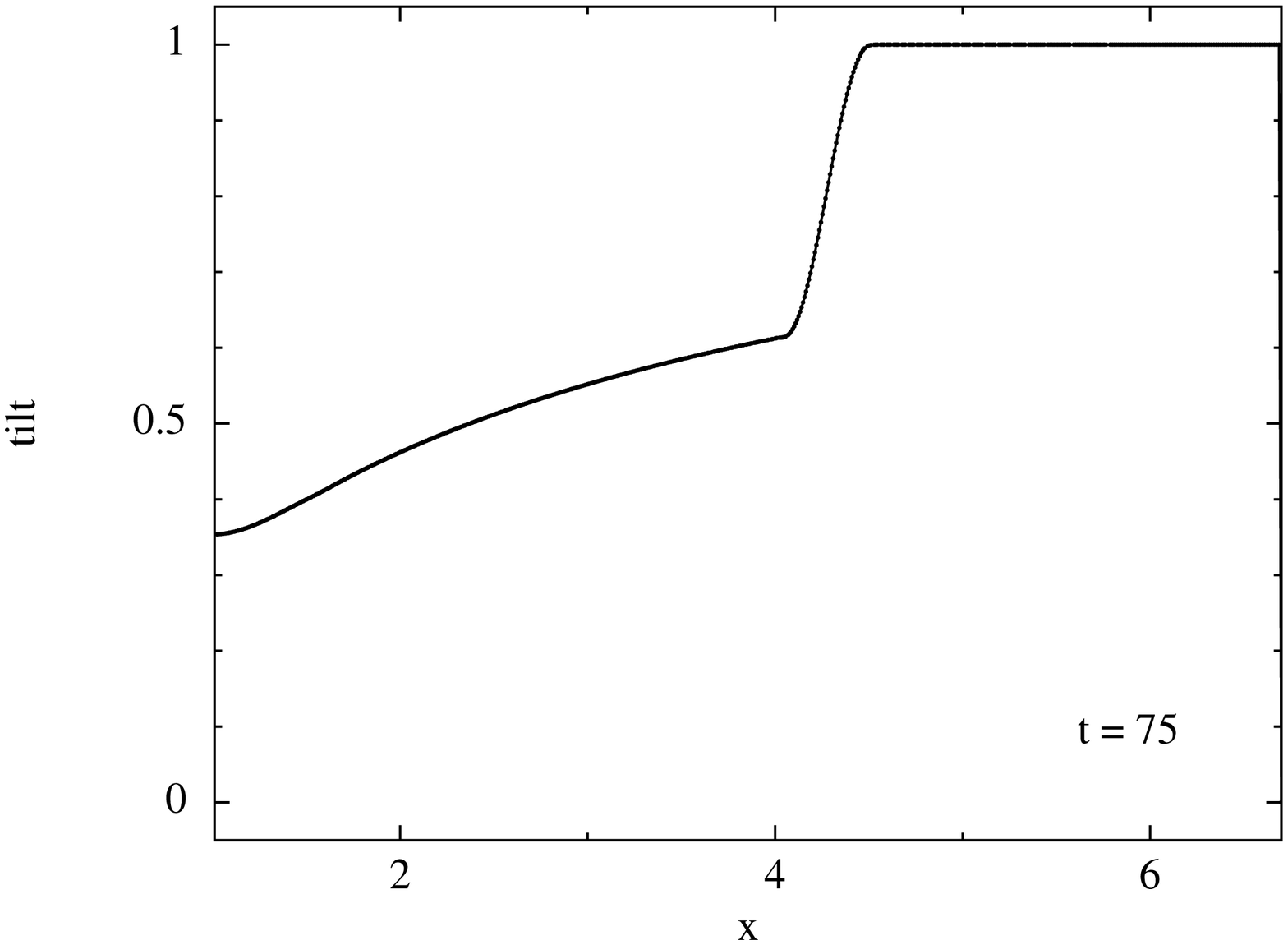}\\
\includegraphics[width=.8\columnwidth]{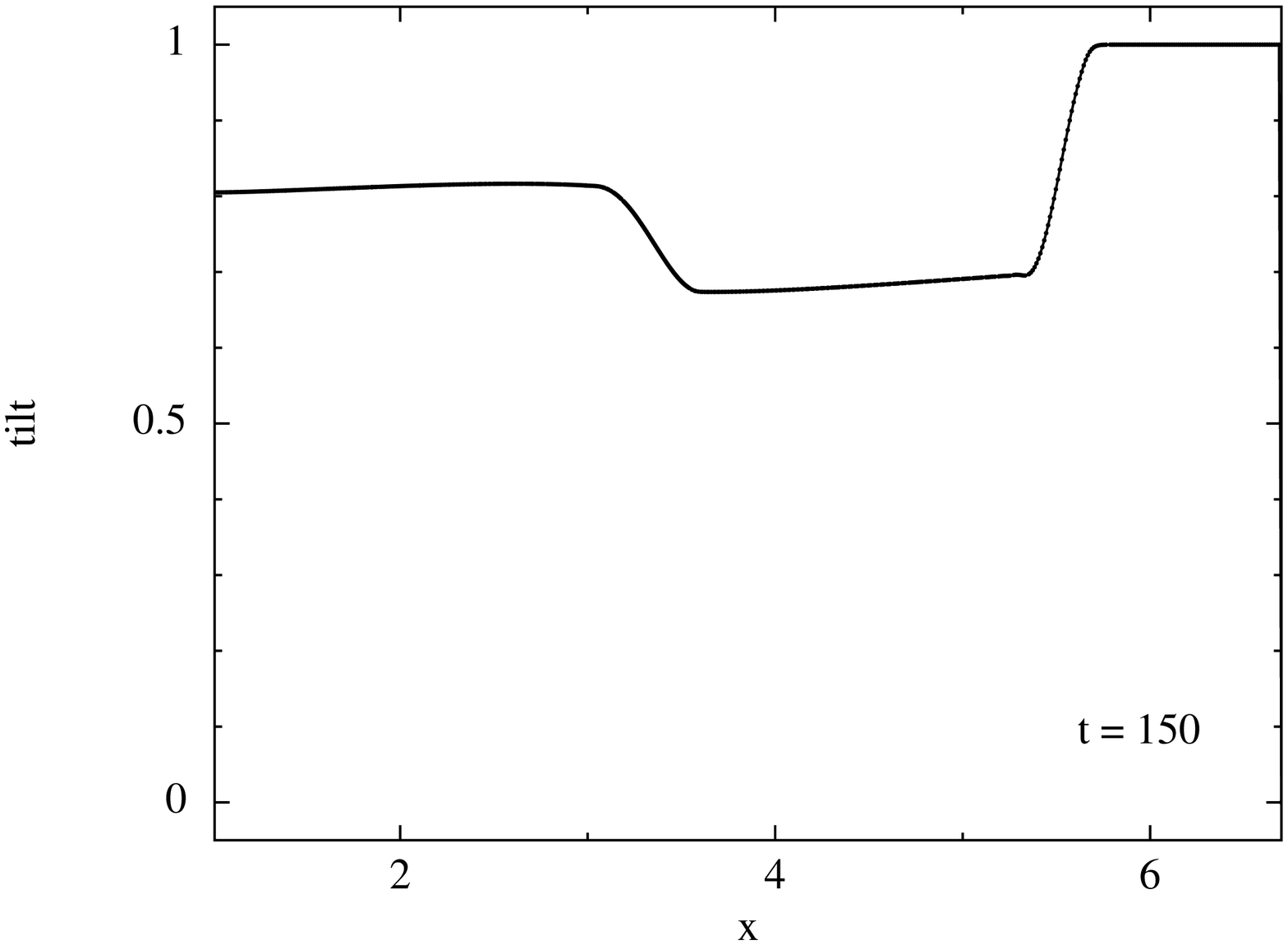}
\includegraphics[width=.8\columnwidth]{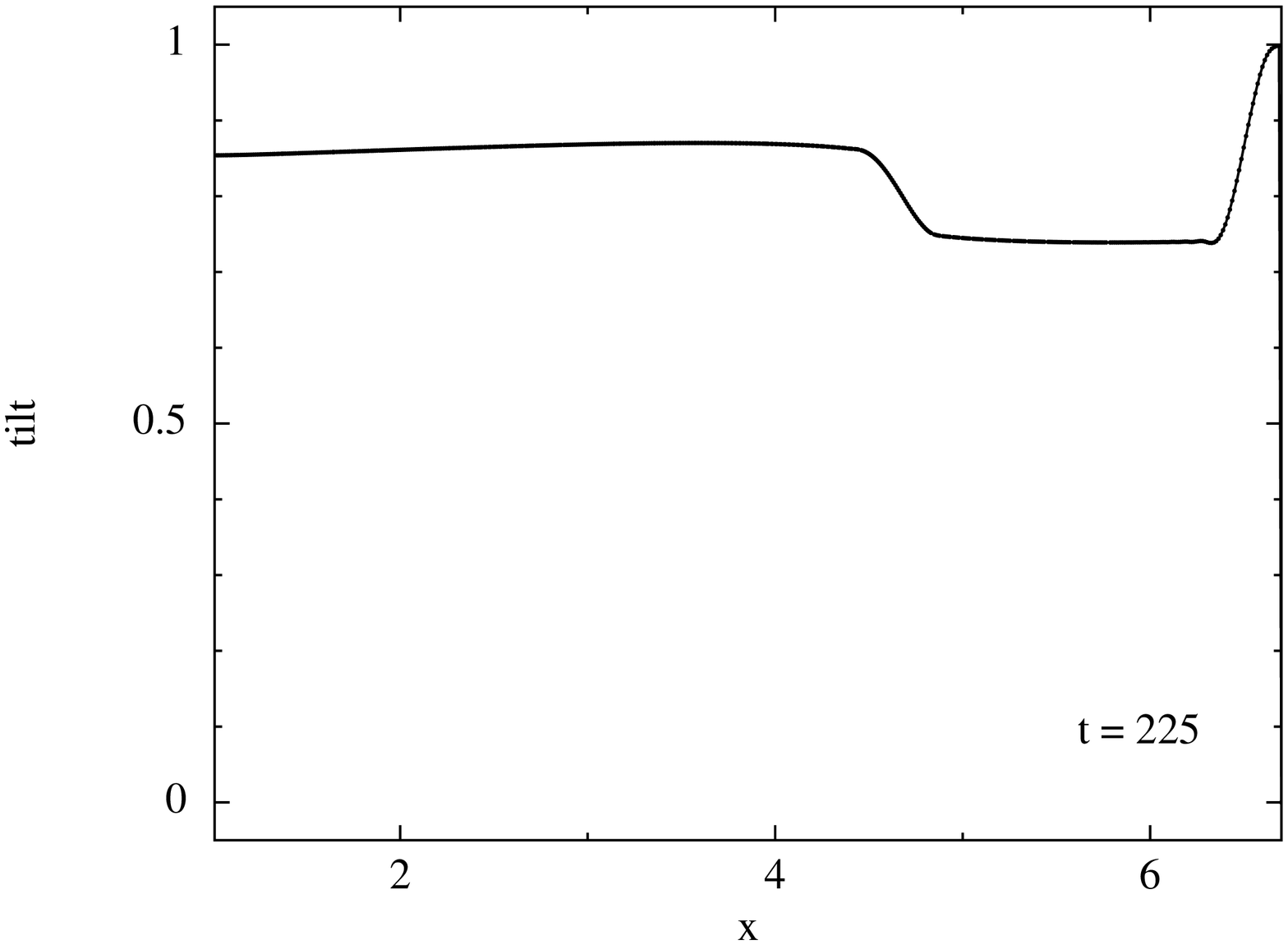}\\
\includegraphics[width=.8\columnwidth]{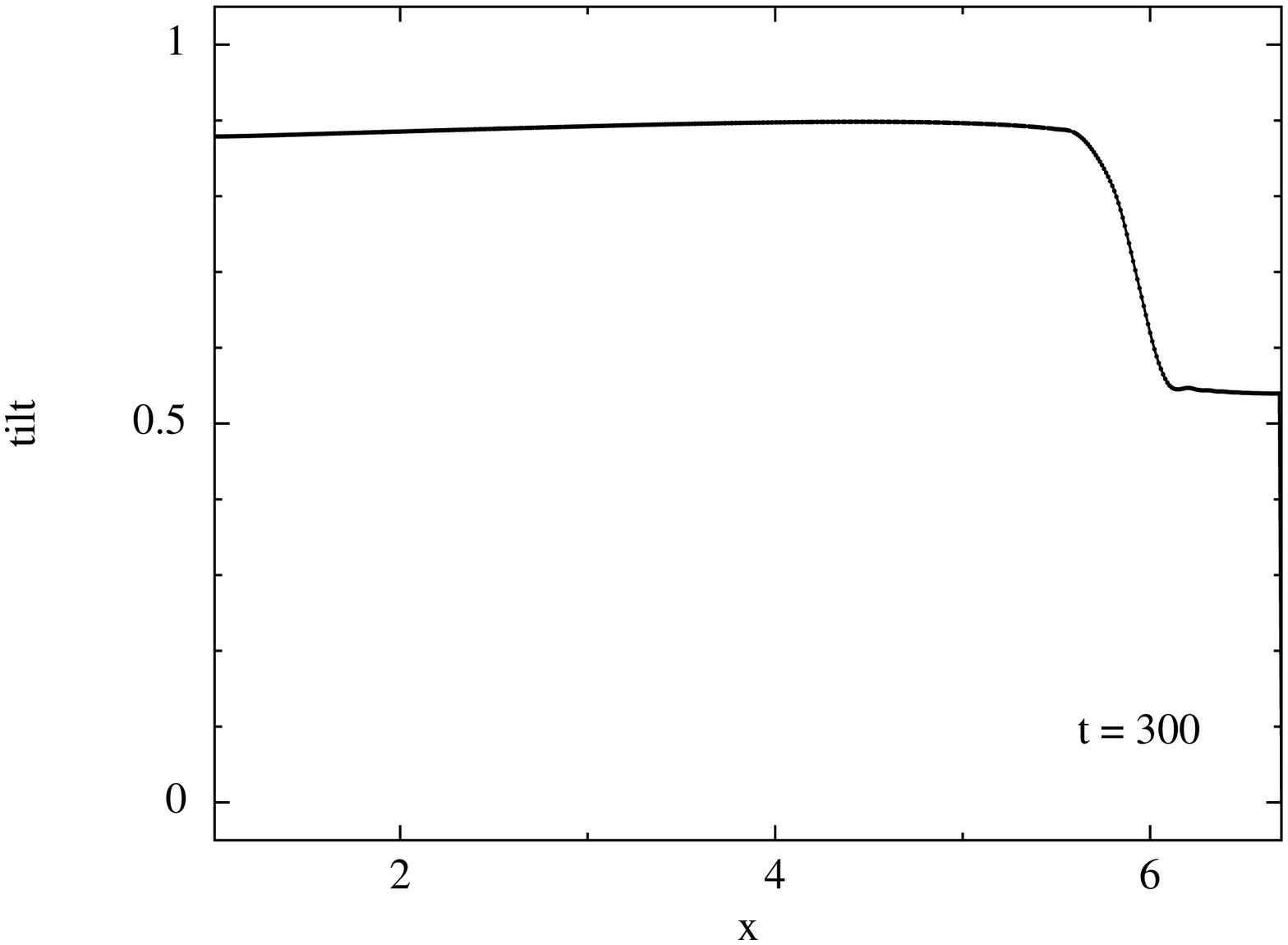}
\includegraphics[width=.8\columnwidth]{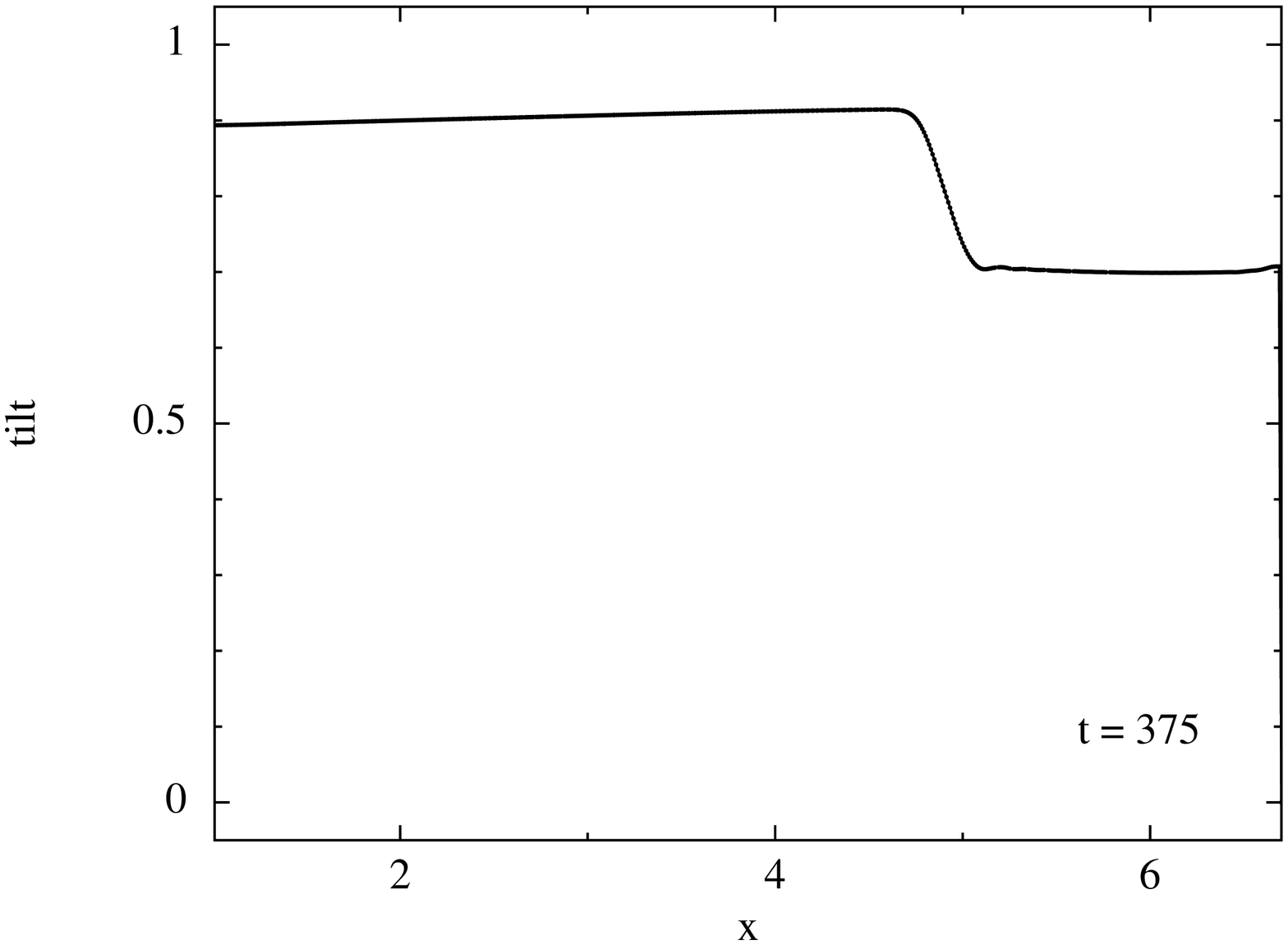}\\
\includegraphics[width=.8\columnwidth]{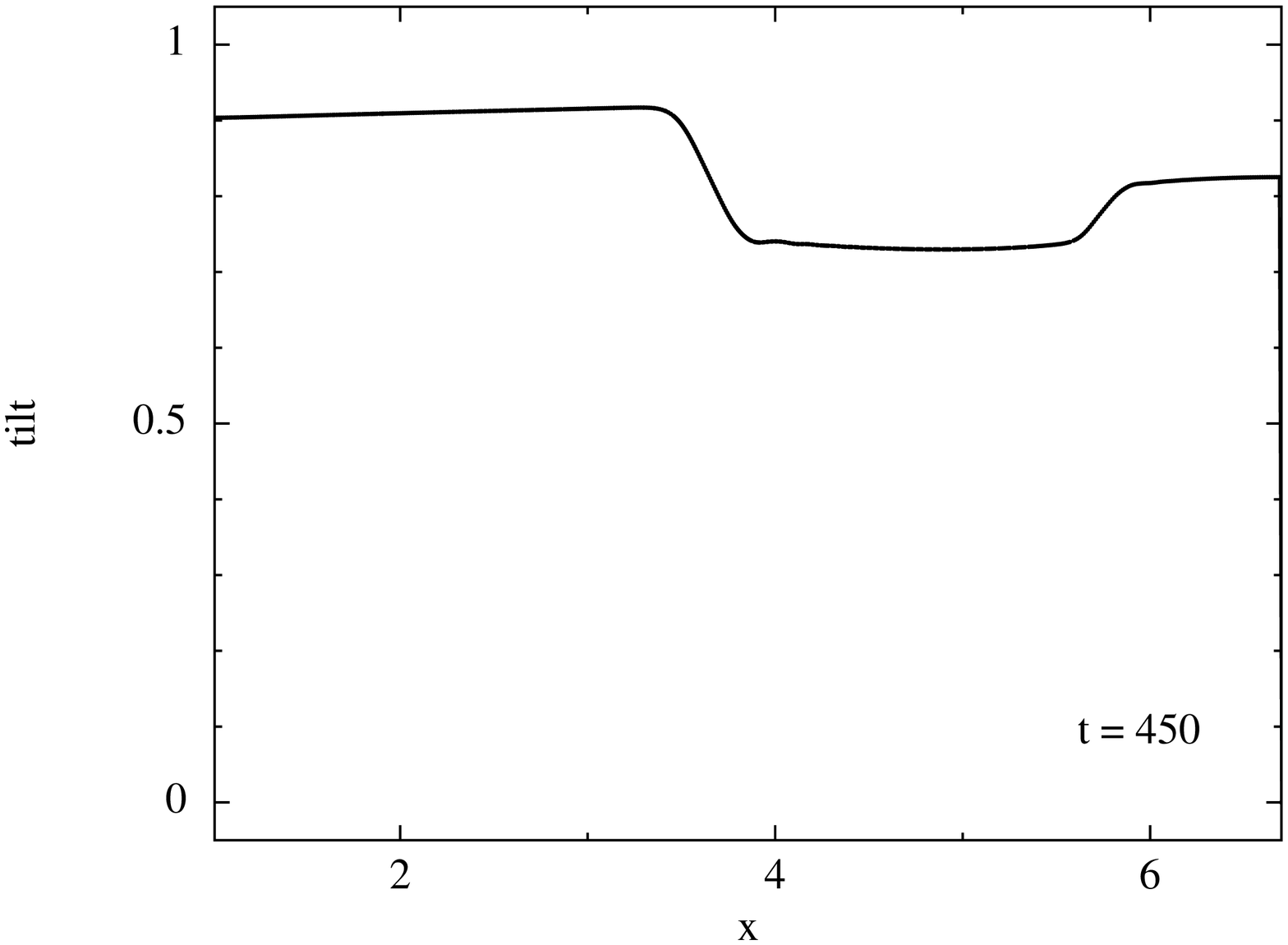}
\includegraphics[width=.8\columnwidth]{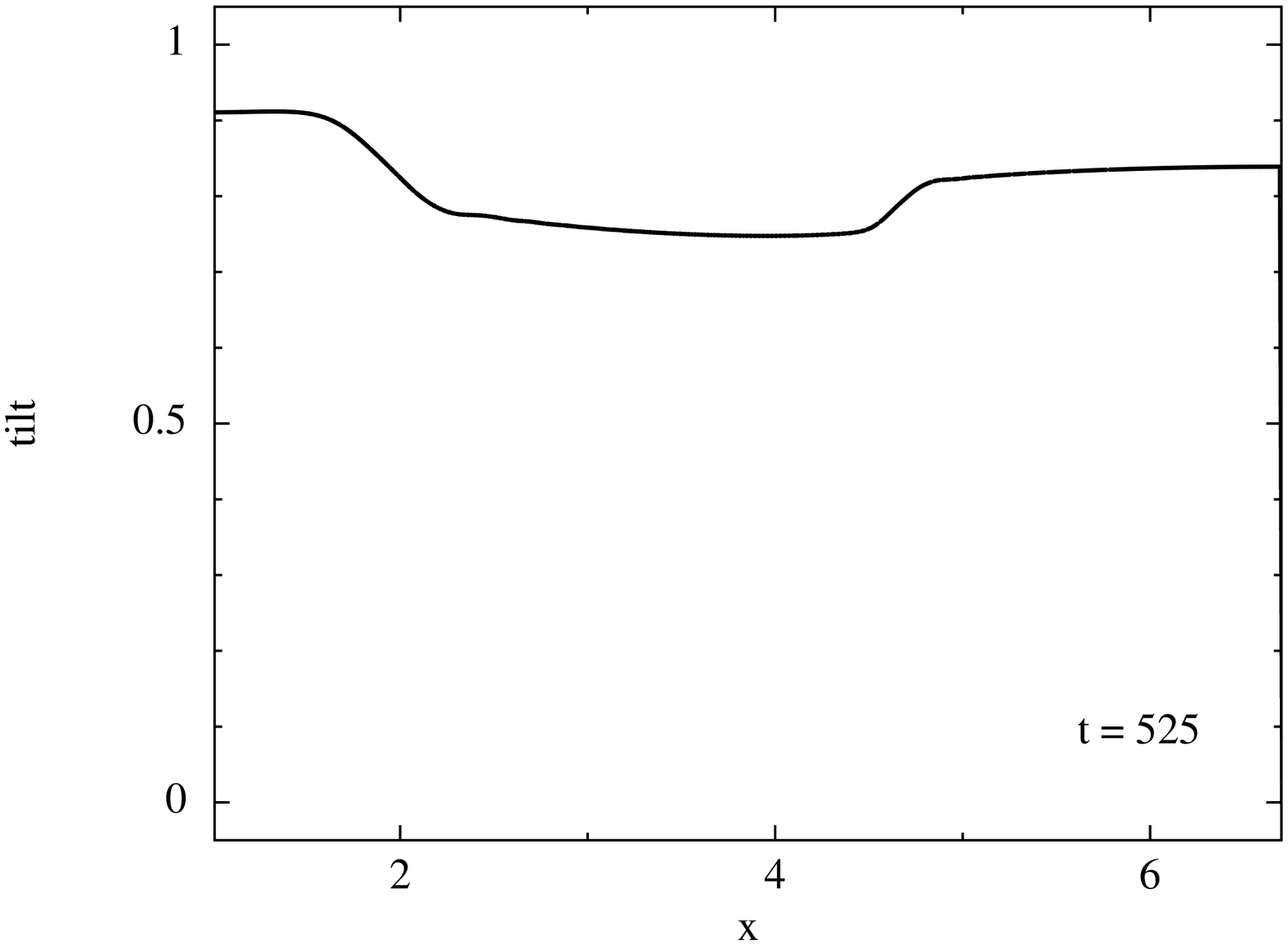}
\end{center}
\caption{From top left to bottom right panel: KH 15D's tilt evolution of an initial warped untwisted disc. In this case, the disc extents from $R_{\rm in}=1$ Au to $R_{\rm out}=6.71$ AU ($p=0.5$). The time values are expressed in dynamical time units $\Omega_0^{-1}$, where $\Omega_0=\Omega{(R=1)}$.}
\label{fig:wave}
\end{figure*}

\begin{figure*}
\begin{center}
\includegraphics[width=.8\columnwidth]{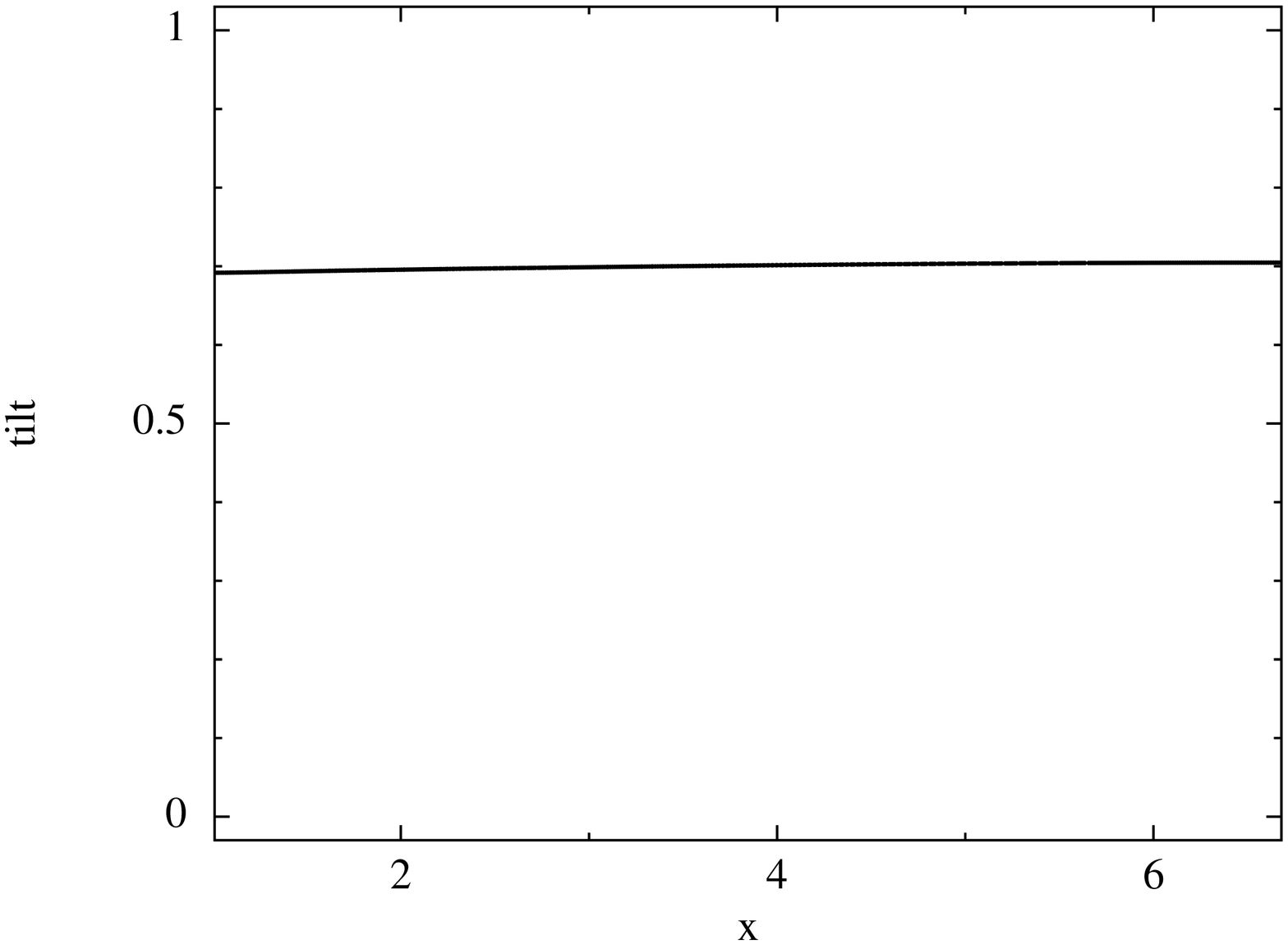}
\includegraphics[width=.8\columnwidth]{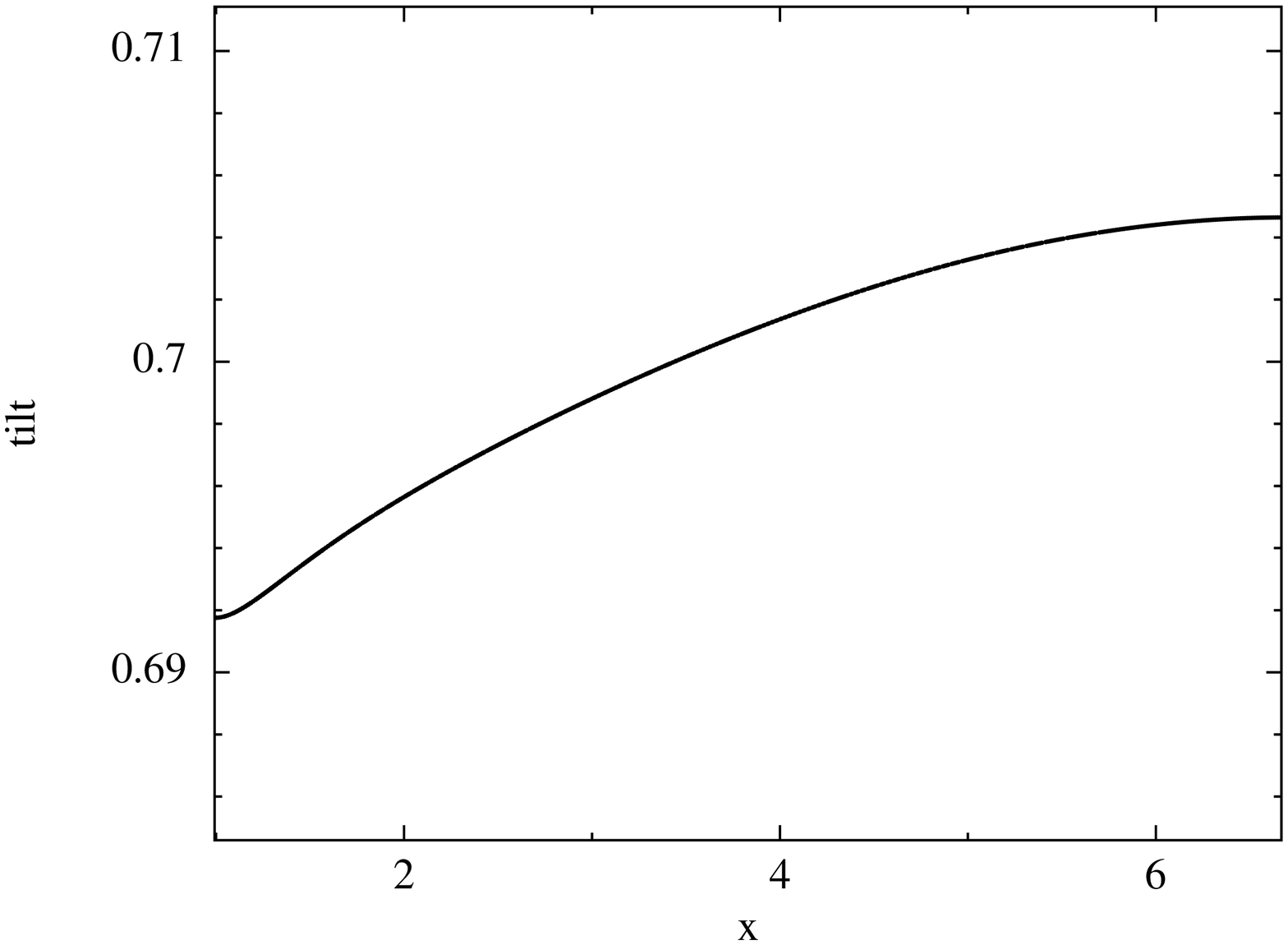}
\end{center}
\caption{Left panel: steady state of KH 15D's tilt reached after $t\approx 4000$ $\Omega_0^{-1}$ for the $p=0.5$ case. Right panel: zoom of the warp. The initial condition of this simulation is the one reported in the top left panel of Fig. \ref{fig:wave}. Note that the fractional variation in inclination is very small, because the external torque is not strong enough to induce a large warp (cf. a similar result by \citealt{foucart_lai_13}, who predict, for the same parameters, a warp larger by roughly a factor 3).}
\label{fig:steady}
\end{figure*}

We run simulations for the two models reported in Table \ref{tab:kh15d}. We set $\alpha=0.05$, a typical value for protostellar discs that ensures that the warp propagates as a bending wave. We have also performed simulations with a different viscosity, in order to explore the dependence of the alignment timescale on $\alpha$. We also adopt the following choice for the remaining parameters: $H_{\rm in}/R_{\rm in}=0.1$, $c_{\rm s}\propto R^{-3/4}$, $\eta=M_KM_K'/M=1/4$ and $a=0.26$AU.

\begin{figure}
\begin{center}
\includegraphics[width=.95\columnwidth]{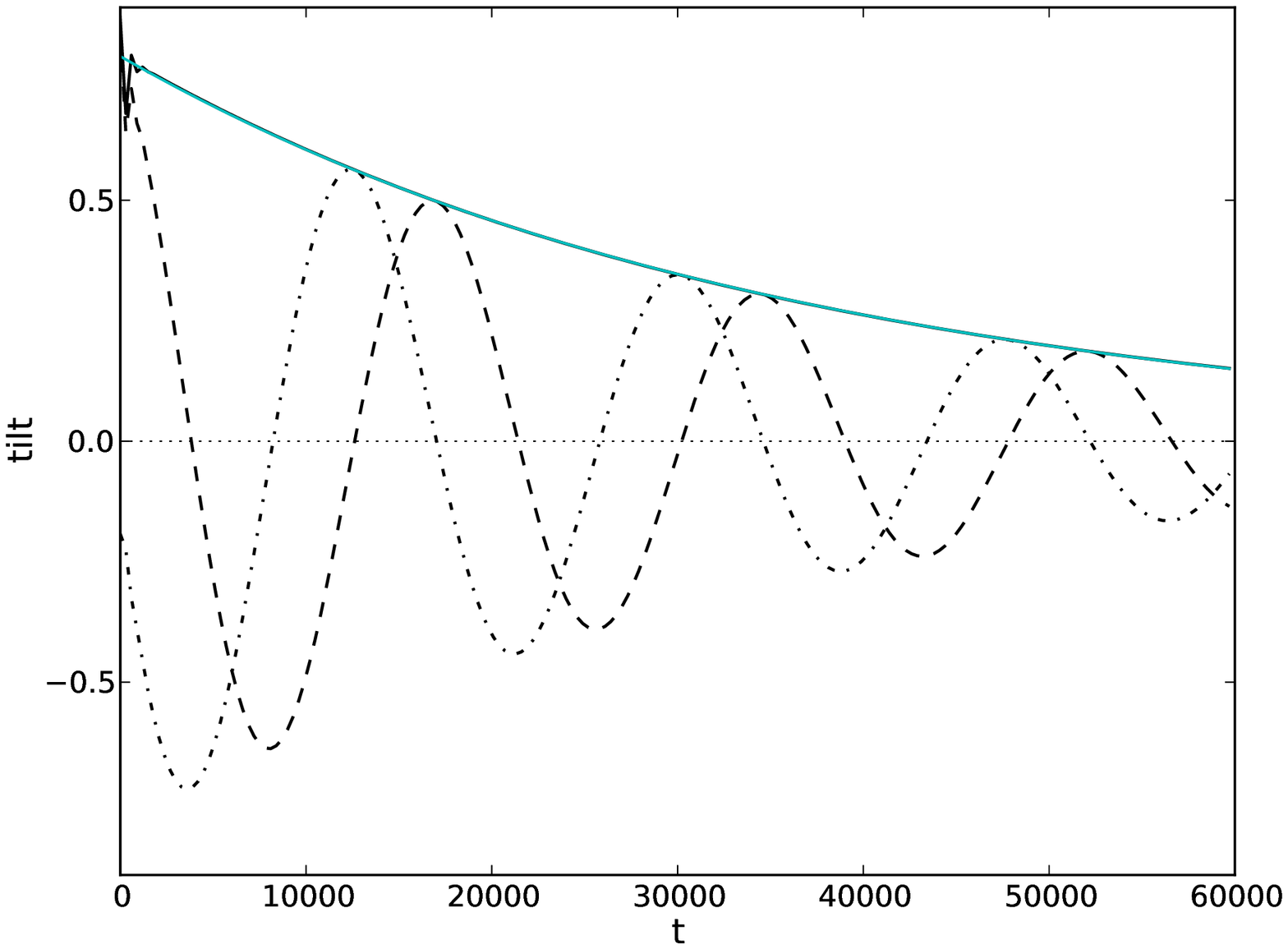}
\includegraphics[width=.95\columnwidth]{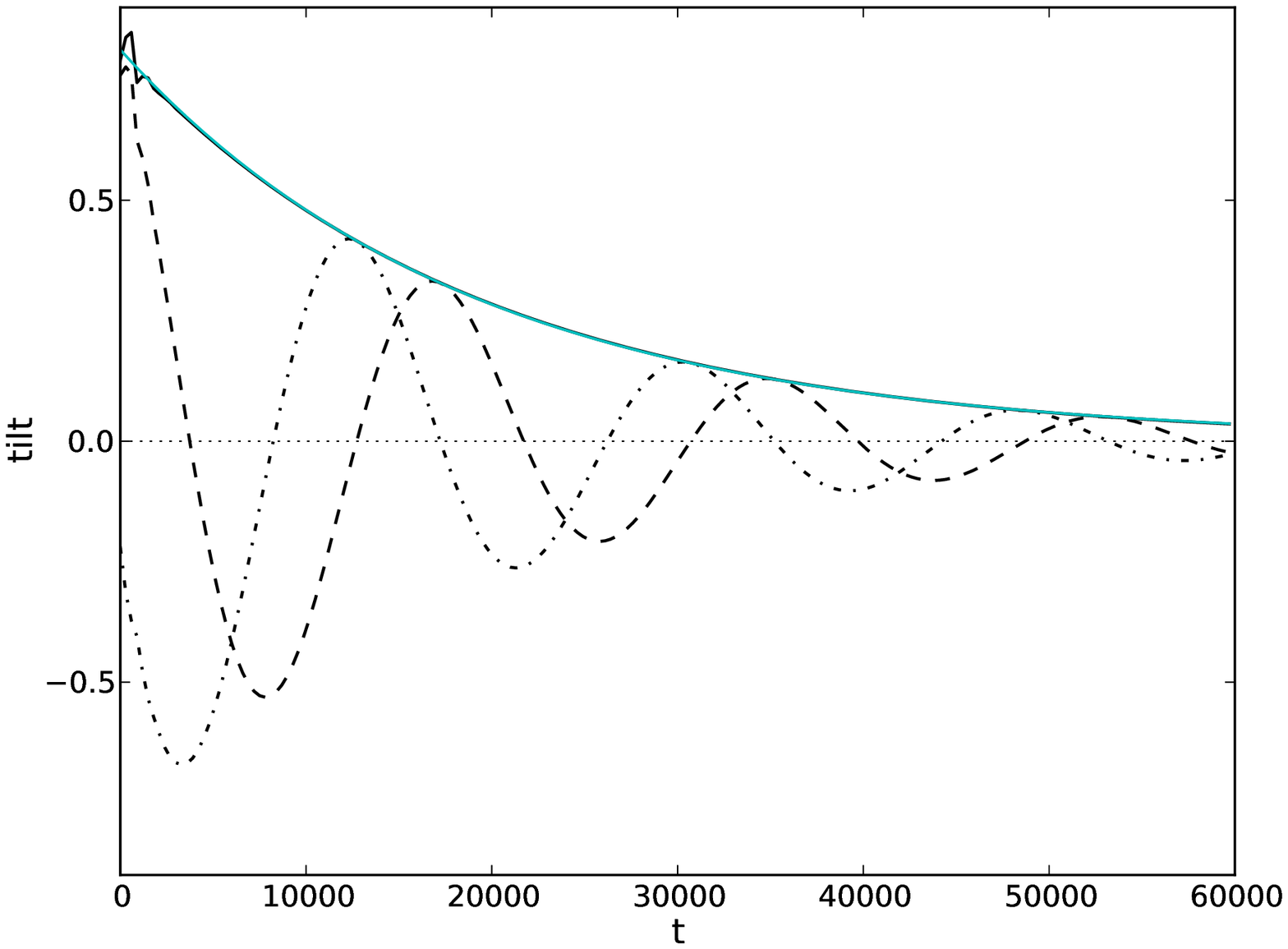}
\end{center}
\caption{Top panel: evolution of the $x$ (dashed line) and $y$ (dash-dotted line) components of the tilt vector at the disc's inner edge for $p=0.5$. Bottom panel: same results for $p=1$. The sinusoidal dependence shows that the two discs are precessing, with an equal period. The black line shows the tilt of the disc. Dissipation (and therefore alignment) is apparent in both the plots. The blue line indicates the exponential decaying fit of the tilt. Finally, note the very early stage of the tilt evolution, when the discs reach a quasi steady state. The time values are expressed in dynamical time units $\Omega_0^{-1}$, where $\Omega_0=\Omega(R=1)$.}
\label{fig:precess}
\end{figure}

We assume a zero-torque boundary condition for the horizontal torques both at the inner and at the outer radius. As initial condition we set a warped untwisted disc. The warp initially propagates inwards and outwards in a wave-like fashion (Fig. \ref{fig:wave}). When the two waves reach the edges of the disc, they bounce back into the disc. This process continues until the tilt reaches a quasi steady state, which corresponds to a stationary wave. The disc tends to an almost flat, misaligned steady state (shown in Fig. 2), with almost no twist.  From Paper I, we expect a small twist to be present for a viscous disc, and as predicted its amplitude increases with the value of $\alpha$. The disc gets to its steady state in a time $\sim4000\ \Omega_0^{-1}$, after which it starts precessing regularly. If we observe the $x$ and $y$ components of the tilt, we observe a sinusoidal dependence on time (Fig.  \ref{fig:precess}). The disc dynamics, the precessional period and the tilt final state do not depend on the initial shape of the warp. We have run other simulations with different initial conditions (a flat misaligned disc and a warped disc with inclination decreasing outwards), and we obtain the same results. We conclude that the disc does precess as a rigid body, with an almost flat, untwisted shape. This result confirms the analytical considerations we have made in section \ref{sec:analyt}. From Fig. \ref{fig:precess} we also notice clearly that the tilt declines exponentially with time, under the effect of viscosity. We discuss this process more in detail below.

From the simulations we obtain a precessional period that is in a very good agreement with the theoretical prediction in both models (see Table \ref{tab:kh15d}).

\begin{table}
\centering
\begin{tabular}{lcc}
\hline
Parameter                      & $p=0.5$ & $p=1$ \\
\hline
$R_{\rm in}$ (AU)              & 1.00    & 1.00  \\
$R_{\rm out}$ (AU)             & 6.71    & 9.00  \\
$T_p$ (yr) analytical          & 2770    & 2770  \\
$T_p$ (yr) numerical           & 2670    & 2680  \\
$\Delta I/\bar{I}$             & 0.012 & 0.040 \\
$t_{\rm align}$ (yr) analytical          & 14051    & 9045  \\
$t_{\rm align}$ (yr) numerical           & 4744     & 3012  \\
\hline
\end{tabular}
\caption{Estimates of the most relevant quantities for the two models. Note the good agreement between the numerical values and the theoretical prediction of $T_p$ and the very small values of the warp amplitude. The alignment timescale estimated via the simulations is of the same order of the theoretical prediction of equation \ref{eq:kh15d_align_time}. The values of $t_{\rm align}$ have been evaluated for $\alpha=0.05$, both analytically and numerically.}
\label{tab:kh15d}
\end{table}

With the numerical simulations, we are able to address another relevant issue: the amplitude of the warp. We know that a warp has to be present, in order for internal pressure torques to compensate the differential external torque generated by the central binary. As shown in Fig. \ref{fig:steady}, however the amplitude of the warp is very small in both models and we report them in Table \ref{tab:kh15d} in terms of $\Delta I/\bar{I}$, where $\Delta I=I(R_{\rm out}) - I(R_{\rm in})$. This is due to the fact that the external torque is not strong enough to induce a large warp. Obviously, the closer is $R_{\rm in}$ to the stars, the larger is the warp. Our amplitudes are significantly smaller than the ones predicted by \citetalias{CM04}, where they deduce the value of $\Delta I/\bar{I}$ from an empirical relation (cf. equation 4 in their paper). Moreover, they propose that the inclination scales with radius as $I\propto R^{-2}$ in a disc dominated by a thermal pressure. This is unphysical. The inclination must increase with radius, since the external torque is much stronger at the inner edge. Finally, we have found that the amplitude of the warp depends very weakly on $\alpha$. The main difference when changing $\alpha$ is the timescale in which the tilt reaches a steady state: when $\alpha$ is low, it takes a longer time to reach it, as expected. The expected amplitude of the warp for discs around circumbinary discs has been recently computed also by \citet{foucart_lai_13}. Our results are in broad agreement with theirs. Applying their estimates to the case at hand, we get $\Delta I/\bar{I}\approx 0.03 (0.02)$ for the case $p=0.5(1)$. While the order of magnitude is the same as that obtained in our simulations, their precise values differ by a factor 2-3, and we do not see the strong dependence on $\alpha$ predicted by their results. Moreover, the dependence on the parameter $p$ appears to be different (which might be due to the fact that we impose a specific profile for the sound speed, while they assume it to depend on $p$ as well). Finally, note that the calculations by \citet{foucart_lai_13} apply to a steady disc extending to infinity, while our ring is narrow radially and is not steady, but rigidly precessing.

We have answered the first two questions listed at the end of section \ref{sec:intro}. We now address the third issue: how long is the alignment timescale $t_{\rm align}$ with respect to the precession period of such a disc? We have evaluated the alignment time by fitting the exponential decay of the tilt over a long time (see Fig. \ref{fig:precess}) with a function of the form $\exp{(-t/t_{\rm align})}$. The fit is very accurate and we obtain $t_{\rm align}=4474$ and $3012$ yr (for $p=0.5$ and $p=1$, respectively). Thus, with $\alpha=0.05$ (the value assumed in these simulations), the disc aligns with the binary within $1-2$ precessional periods. However, we have verified that $t_{\rm align}$ is inversely proportional to $\alpha$, as predicted by equation \ref{eq:kh15d_align_time}. For example, if we consider the model with $p=1$, we can express the alignment timescale as a function of $\alpha$ only:
\begin{equation}
t_{\rm align} \approx 3000 \left(\frac{0.05}{\alpha}\right)\ {\rm yr}.
\end{equation}
In order to maintain a non-zero inclination for a significant number of precessional times, we require $\alpha\lesssim 0.005$, a relatively low value, that may be compatible with a rather evolved protostellar disc, that is expected to be cold such that the magneto-rotational instability might not provide a large stress.

Note that our numerically derived values of $t_{\rm align}$ are of the same order of those estimated through equation \ref{eq:kh15d_align_time}, from which we would obtain $t_{\rm align}=14051$ and $9045$ yr (for $p=0.5$ and $p=1$, respectively) when $\alpha=0.05$, a factor three larger than what measured from the simulations.

\subsection{Chiang and Murray-Clay model}
\label{sec:CM_par}

 In this paper most of the work is based on the observational model by \citetalias{CM04}. Given the parameters values known from the observations (reported at the end of section \ref{sec:obs}), \citetalias{CM04} develop two models for the disc extent. A first one, where the differential precession is prevented by internal pressure, and a second one, where it is prevented by the disc self-gravity. In this section, we briefly compare the first of these two models with the ones developed in this paper. \citetalias{CM04} assume a disc inner edge of $2.25$ AU, and an outer edge of $3.75$ AU. They use these two parameters in order to have a precessional period of $\sim 2770$ yr (cf. equation 1 of their paper) and a narrow disc. Note that the choice of $R_{\rm in}=2.25$ AU is arbitrary. We have performed numerical simulations with such parameters for the disc, and we obtain a precessional period of $3300$ yr (which is also confirmed by the analytical formula of equation \ref{eq:precper}). Thus, also in this case the precessional period is in rough agreement with what required observationally, although in order to have such a narrow width, one is forced to fix the inner edge of the disc further out than the expected truncation radius due to the inner eccentric binary. Such a narrow disc model, on the other hand, has the property of decaying much more slowly than an extended one, thus not requiring a very low viscosity in order to be maintained for long enough. In particular, we measure numerically a decay time of the order of $10^6$ years. 
 
 Finally, we also measure the warp amplitude needed to balance the binary torques and find that with these parameters it is as small as $\Delta I/\bar{I}\approx 10^{-4}$, two orders of magnitude smaller than that estimated by CM.

\section{Conclusions}
\label{sec:concl}

In this paper, we have developed the precessing disc model, introduced by CM, to describe the peculiar eclipsing binary system KH 15D. In particular, we have improved on previous analytical estimates of the disc fundamental parameters, such as its radial extent and the magnitude of the warp. We assume, following CM, that the light curve of the system is well described by a model where a stellar binary is surrounded by an inclined disc, rigidly precessing with a period of about 2770 years. We fix the inner edge of the disc to the expected tidal truncation radius due to the binary, at $\approx 1$ AU and find that, in order to reproduce the required precession timescale, the disc needs to be wider than previously assumed, extending out to 5-10 AU, depending on the slope of the disc surface density profile. 

We test our model predictions by using the bending wave time dependent formalism introduced by \citet{lubow_ogilvie2000}. We find indeed that the disc rapidly attains a rigidly precessing shape, with precession timescales matching our analytical predictions. In such a steady precessing configuration, the disc needs to be warped, in order to counteract the binary torques. However, the warp amplitude $\Delta I/\bar{I}$ across the disc radial extent is much smaller than the one predicted by the formulae by CM for the corresponding parameters, and is of the order of $\sim 0.01$. Note that CM had incorrectly attributed a ``reverse'' warp (where the disc inclination with respect to the binary \emph{decreases} outwards) in the case of a warp propagating through pressure, rather than via self-gravity. Here we demonstrate that also in this case the disc is more aligned with the binary in its inner parts and less in the outer parts, in agreement with what indicated by recent observations \citep{capelo12}.

With such a wider disc, the issue of its long term maintenance arises, as wider discs tend to align more rapidly with the binary plane. Our time dependent calculations, that also include a dissipative viscous term, allow us to compute such decay rate. We find that the analytical formula for the decay rate derived by \citet{bate00} in the case of a tidally confined circumprimary disc works well also in our case. If $\alpha=0.05$, the disc aligns with the binary within 1-2 precessional periods, depending on the model parameters. However, we have verified that the decay rate is inversely proportional to $\alpha$, so that with $\alpha \lesssim 0.005$ (which might be not unreasonable for a cold and evolved protostellar disc such as the one surrounding KH 15D) a non zero inclination can be maintained for a reasonable time. If the decay rate could be measured through its effect on the long term light curve of the system, this would give a constraint on the magnitude of $\alpha$. This kind of modelling shows that one can use warped discs effectively in order to put interesting constraints on the magnitude of the disc viscosity based on observed systems \citep[see also][]{king13}.

We have also tested numerically the model parameters introduced by CM. Such models are all characterised by a larger inner radius of the disc with respect to what expected based on the tidal truncation from the binary torques, and by a smaller radial extent with respect to ours. Nevertheless, also such models do reproduce the correct precession period, but we find that the required warp amplitude is much smaller (by two orders of magnitude) than what inferred by CM. Such marrow rings, on the other hand, are able to maintain a non-zero inclination for much longer than our extended discs. 

Our model is still simplified and approximate in many respects. First of all, note that the precessional time scale of 2770 year is not uniquely fixed by the observations, and might well range between 2000 and 3000 years. This would reflect in a relatively modest change of our inferred parameters, such that the outer radius of the disc might vary between $5.68$ and $6.99$ AU (in the $p=0.5$ case). 

The most important limitation of our model is the assumption that the binary orbit is circular, while we know that its eccentricity should range over $0.68<e<0.80$. The effects of an eccentric orbit might be to push further back the inner edge of the disc. A detailed examination of such effects would require full 3D numerical simulations. 

Finally, as in the case of the models by CM, an outstanding issue is the origin of such a narrow disc. Most likely the circumbinary disc is being confined by a third body (such a planetary companion or a very low mass star). Also the effects of such purported body on the long term evolution of the system needs to be explored more thoroughly.

\section*{Acknowledgements}
We thank Gordon Ogilvie and Jim Pringle for useful advice and an anonymous referee for a careful reading of the manuscript. SF thanks the Science and Technology Facility Council and the Isaac Newton Trust for the award of a studentship. Figs. \ref{fig:wave} and \ref{fig:steady} were produced using \textsc{SPLASH} \citep{price07}.

\appendix

\bibliography{warp_bib2}

\begin{thebibliography}{}

\bibitem[\protect\citeauthoryear{{Agol}, {Barth}, {Wolf} \& {et al.}}{{Agol}
  et~al.}{2004}]{agol2004}
{Agol} E.,  {Barth} A.~J.,  {Wolf} S.,    {et al.} 2004, \apj, 600, 781

\bibitem[\protect\citeauthoryear{{Akeson}, {Rice}, {Boden} \& {et
  al.}}{{Akeson} et~al.}{2007}]{akeson07}
{Akeson} R.~L.,  {Rice} W.~K.~M.,  {Boden} A.~F.,    {et al.} 2007, \apj, 670,
  1240

\bibitem[\protect\citeauthoryear{{Artymowicz} \& {Lubow}}{{Artymowicz} \&
  {Lubow}}{1994}]{art_lubow94}
{Artymowicz} P.,  {Lubow} S.~H.,  1994, \apj, 421, 651

\bibitem[\protect\citeauthoryear{{Bate}, {Bonnell}, {Clarke} \& {et
  al.}}{{Bate} et~al.}{2000}]{bate00}
{Bate} M.~R.,  {Bonnell} I.~A.,  {Clarke} C.~J.,    {et al.} 2000, \mnras, 317,
  773

\bibitem[\protect\citeauthoryear{{Bate}, {Lodato} \& {Pringle}}{{Bate}
  et~al.}{2010}]{bate10}
{Bate} M.~R.,  {Lodato} G.,    {Pringle} J.~E.,  2010, \mnras, 401, 1505

\bibitem[\protect\citeauthoryear{{Beust} \& {Dutrey}}{{Beust} \&
  {Dutrey}}{2005}]{beust_dutrey05}
{Beust} H.,  {Dutrey} A.,  2005, \aap, 439, 585

\bibitem[\protect\citeauthoryear{{Bonnell}, {Arcoragi}, {Martel} \& {et
  al.}}{{Bonnell} et~al.}{1992}]{bonnell92}
{Bonnell} I.,  {Arcoragi} J.-P.,  {Martel} H.,    {et al.} 1992, \apj, 400, 579

\bibitem[\protect\citeauthoryear{{Capelo}, {Herbst}, {Leggett} \& {et
  al.}}{{Capelo} et~al.}{2012}]{capelo12}
{Capelo} H.~L.,  {Herbst} W.,  {Leggett} S.~K.,    {et al.} 2012, \apjl, 757,
  L18

\bibitem[\protect\citeauthoryear{{Chiang} \& {Murray-Clay}}{{Chiang} \&
  {Murray-Clay}}{2004}]{CM04}
{Chiang} E.~I.,  {Murray-Clay} R.~A.,  2004, \apj, 607, 913

\bibitem[\protect\citeauthoryear{{Clarke}, {Bonnell} \& {Hillenbrand}}{{Clarke}
  et~al.}{2000}]{clarke00}
{Clarke} C.~J.,  {Bonnell} I.~A.,    {Hillenbrand} L.~A.,  2000, Protostars and
  Planets IV, p.~151

\bibitem[\protect\citeauthoryear{{Dutrey}, {Guilloteau} \& {Simon}}{{Dutrey}
  et~al.}{1994}]{dutrey94}
{Dutrey} A.,  {Guilloteau} S.,    {Simon} M.,  1994, \aap, 286, 149

\bibitem[\protect\citeauthoryear{{Facchini}, {Lodato} \& {Price}}{{Facchini}
  et~al.}{2013}]{facchini13_1}
{Facchini} S.,  {Lodato} G.,    {Price} D.~J.,  2013, \mnras, accepted for
  publication

\bibitem[\protect\citeauthoryear{{Foucart} \& {Lai}}{{Foucart} \&
  {Lai}}{2013}]{foucart_lai_13}
{Foucart} F.,  {Lai} D.,  2013, \apj, 764, 106

\bibitem[\protect\citeauthoryear{{Ghez}, {Neugebauer} \& {Matthews}}{{Ghez}
  et~al.}{1993}]{ghez93}
{Ghez} A.~M.,  {Neugebauer} G.,    {Matthews} K.,  1993, \aj, 106, 2005

\bibitem[\protect\citeauthoryear{{Heller}}{{Heller}}{1995}]{heller95}
{Heller} C.~H.,  1995, \apj, 455, 252

\bibitem[\protect\citeauthoryear{{Herbst}, {Hamilton}, {Vrba} \& {et
  al.}}{{Herbst} et~al.}{2002}]{herbst2002}
{Herbst} W.,  {Hamilton} C.~M.,  {Vrba} F.~J.,    {et al.} 2002, \pasp, 114,
  1167

\bibitem[\protect\citeauthoryear{{Herbst}, {LeDuc}, {Hamilton} \& {et
  al.}}{{Herbst} et~al.}{2010}]{herbst2010}
{Herbst} W.,  {LeDuc} K.,  {Hamilton} C.~M.,    {et al.} 2010, \aj, 140, 2025

\bibitem[\protect\citeauthoryear{{Johnson}, {Marcy}, {Hamilton} \& {et
  al.}}{{Johnson} et~al.}{2004}]{johnson2004}
{Johnson} J.~A.,  {Marcy} G.~W.,  {Hamilton} C.~M.,    {et al.} 2004, \aj, 128,
  1265

\bibitem[\protect\citeauthoryear{{Johnson} \& {Winn}}{{Johnson} \&
  {Winn}}{2004}]{johnson_winn2004}
{Johnson} J.~A.,  {Winn} J.~N.,  2004, \aj, 127, 2344

\bibitem[\protect\citeauthoryear{{Kearns} \& {Herbst}}{{Kearns} \&
  {Herbst}}{1998}]{kearns1998}
{Kearns} K.~E.,  {Herbst} W.,  1998, \aj, 116, 261

\bibitem[\protect\citeauthoryear{{King}, {Livio}, {Lubow} \& {Pringle}}{{King}
  et~al.}{2013}]{king13}
{King} A.~R.,  {Livio} M.,  {Lubow} S.~H.,    {Pringle} J.~E.,  2013, \mnras,
  431, 2655

\bibitem[\protect\citeauthoryear{{Larwood}, {Nelson}, {Papaloizou} \&
  {Terquem}}{{Larwood} et~al.}{1996}]{larwood96}
{Larwood} J.~D.,  {Nelson} R.~P.,  {Papaloizou} J.~C.~B.,    {Terquem} C.,
  1996, \mnras, 282, 597

\bibitem[\protect\citeauthoryear{{Larwood} \& {Papaloizou}}{{Larwood} \&
  {Papaloizou}}{1997}]{larwood_pap97}
{Larwood} J.~D.,  {Papaloizou} J.~C.~B.,  1997, \mnras, 285, 288

\bibitem[\protect\citeauthoryear{{Lodato}}{{Lodato}}{2008}]{lodato08}
{Lodato} G.,  2008, \nar, 52, 21

\bibitem[\protect\citeauthoryear{{Lodato} \& {Pringle}}{{Lodato} \&
  {Pringle}}{2006}]{lodato_pringle06}
{Lodato} G.,  {Pringle} J.~E.,  2006, \mnras, 368, 1196

\bibitem[\protect\citeauthoryear{{Lubow} \& {Ogilvie}}{{Lubow} \&
  {Ogilvie}}{2000}]{lubow_ogilvie2000}
{Lubow} S.~H.,  {Ogilvie} G.~I.,  2000, \apj, 538, 326

\bibitem[\protect\citeauthoryear{{Lubow} \& {Ogilvie}}{{Lubow} \&
  {Ogilvie}}{2001}]{lubow_ogilvie01}
{Lubow} S.~H.,  {Ogilvie} G.~I.,  2001, \apj, 560, 997

\bibitem[\protect\citeauthoryear{{Lubow}, {Ogilvie} \& {Pringle}}{{Lubow}
  et~al.}{2002}]{LOP02}
{Lubow} S.~H.,  {Ogilvie} G.~I.,    {Pringle} J.~E.,  2002, \mnras, 337, 706

\bibitem[\protect\citeauthoryear{{Moeckel} \& {Bally}}{{Moeckel} \&
  {Bally}}{2006}]{moeckel06}
{Moeckel} N.,  {Bally} J.,  2006, \apj, 653, 437

\bibitem[\protect\citeauthoryear{{Ogilvie}}{{Ogilvie}}{1999}]{ogilvie99}
{Ogilvie} G.~I.,  1999, \mnras, 304, 557

\bibitem[\protect\citeauthoryear{{Papaloizou} \& {Lin}}{{Papaloizou} \&
  {Lin}}{1995}]{papaloizou_lin95}
{Papaloizou} J.~C.~B.,  {Lin} D.~N.~C.,  1995, \apj, 438, 841

\bibitem[\protect\citeauthoryear{{Papaloizou} \& {Pringle}}{{Papaloizou} \&
  {Pringle}}{1983}]{papaloizou_pringle83}
{Papaloizou} J.~C.~B.,  {Pringle} J.~E.,  1983, \mnras, 202, 1181

\bibitem[\protect\citeauthoryear{{Papaloizou} \& {Terquem}}{{Papaloizou} \&
  {Terquem}}{1995}]{pap_terquem95}
{Papaloizou} J.~C.~B.,  {Terquem} C.,  1995, \mnras, 274, 987

\bibitem[\protect\citeauthoryear{{Price}}{{Price}}{2007}]{price07}
{Price} D.~J.,  2007, \pasa, 24, 159

\bibitem[\protect\citeauthoryear{{Pringle}}{{Pringle}}{1992}]{pringle92}
{Pringle} J.~E.,  1992, \mnras, 258, 811

\bibitem[\protect\citeauthoryear{{Shakura} \& {Sunyaev}}{{Shakura} \&
  {Sunyaev}}{1973}]{shakura73}
{Shakura} N.~I.,  {Sunyaev} R.~A.,  1973, \aap, 24, 337

\bibitem[\protect\citeauthoryear{{Simon}, {Ghez}, {Leinert} \& {et
  al.}}{{Simon} et~al.}{1995}]{simon95}
{Simon} M.,  {Ghez} A.~M.,  {Leinert} C.,    {et al.} 1995, \apj, 443, 625

\bibitem[\protect\citeauthoryear{{Winn}, {Garnavich}, {Stanek} \& {et
  al.}}{{Winn} et~al.}{2003}]{winn2003}
{Winn} J.~N.,  {Garnavich} P.~M.,  {Stanek} K.~Z.,    {et al.} 2003, \apjl,
  593, L121

\bibitem[\protect\citeauthoryear{{Winn}, {Holman}, {Johnson} \& {et
  al.}}{{Winn} et~al.}{2004}]{winn2004}
{Winn} J.~N.,  {Holman} M.~J.,  {Johnson} J.~A.,    {et al.} 2004, \apjl, 603,
  L45

\end{thebibliography}

\label{lastpage}
\end{document}